\documentclass[pra,twocolumn,showpacs,preprintnumbers,amsmath,amssymb,tightenlines]{revtex4}

\usepackage{graphicx}% Include figure files
\usepackage{dcolumn}% Align table columns on decimal point
\usepackage{bm}% bold math
\usepackage{epsfig}
\usepackage{stmaryrd}

\newcommand{\uu}[2]{u_{#1}({\mathbf #2})}
\newcommand{\vv}[2]{v_{#1}({\mathbf #2})}

\newcommand{\ssection}[1]{{\noi  \it #1:}}
\newcommand{\ofx}{(\mathbf{x})}
 \newcommand{\intas}{\int d \mathbf{x}\:}
\newcommand{\comm}[2]{\big[#1,#2\big]}

\newcommand{\tabvspace}{\Big.}
\newcommand{\sub}[2]{{#1}_{\mbox{\!\! \scriptsize #2}}}

\def\noi{\noindent}
\def\beq{\begin{equation}}
\def\eeq{\end{equation}}

\def\CR{\nonumber\\[0.15cm]}
\newcommand{\rref}[1]{Ref.~\cite{#1}}

\newcommand{\eref}[1]{Eq.~(\ref{#1})}

\newcommand{\cref}[1]{chapter~\ref{#1}}

\newcommand{\Cref}[1]{Chapter~\ref{#1}}
\newcommand{\tref}[1]{table~\ref{#1}}

\begin{document}

\title{Phonon background versus analogue Hawking radiation in Bose-Einstein condensates}
\author{S. W\"uster}
\affiliation{Max Planck Institute for the Physics of Complex Systems, N\"othnitzer Strasse 38, 01187 Dresden, Germany}
\email{sew654@pks.mpg.de}

\begin{abstract}
We determine the feasibility of detecting analogue Hawking radiation in a Bose-Einstein condensate in the presence of atom loss induced heating. We find that phonons created by three-body losses overshadow those due to analogue Hawking radiation. To overcome this problem, three-body losses may have to be suppressed, for example as proposed by Search {\it et al.} [Phys.~Rev.~Lett.~{\bf 92} 140401 (2004)]. The reduction of losses to a few percent of their normal rate is typically sufficient to suppress the creation of loss phonons on the time scale of fast analogue Hawking phonon detection. 
\end{abstract}

\maketitle

%%%%%%%%%%%%%%%%%%%%%%%%%%%%%%%%
\ssection{Introduction}
An analogue of the Hawking effect~\cite{hawking:hr1,hawking:hr2} should in principle exist in cold moving fluids~\cite{unruh:bholes}. A fluid whose velocity profile contains a transition from subsonic to supersonic flow will emit a thermal radiation of quantized sound waves (phonons). Gaseous Bose-Einstein condensates (BECs) were often considered most promising for the observation of this phenomenon~\cite{visser:review,visser:towards, garay:prl, garay:pra, giovanazzi:horizon, barcelo:diffmetric,fabbri:wigner}.
However, for accessible Hawking temperatures the condensate must have densities for which three-body losses become relevant~\cite{wuester:horizon}. These strongly constrain the time available for phonon detection already on the level of the mean-field. 
Here, we consider limitations arising from three-body loss in a quantum treatment \cite{schuetzhold:rsoc} and show that they are even more severe. 

The primary consequence of three-body loss is a reduction of the condensate density in time, resulting in a decrease of the Hawking temperature. Three-body losses also heat the condensate by driving the many-body quantum state away from the Bogoliubov vacuum~\cite{dziarmaga:lossheating}. This creates phonons, which contribute in general to a background indistinguishable from analogue Hawking radiation~\cite{footnote:correlations}. Here we compare these two effects and show that in equilibrium the loss-phonons always overshadow those created by the Hawking effect. For the densities required to observe analogue Hawking radiation~\cite{wuester:horizon} the loss-phonons are created on the same time-scale as the Hawking phonons.

Our findings indicate that it may be necessary to suppress three-body losses in a BEC in order to observe analogue Hawking radiation. Fortunately, suppression schemes exist~\cite{search:lossinhibit, yurovsky:lessloss}. A suppression will have a two-fold benefit: Firstly it increases the time-scale for the loss-heating to reach equilibrium, making it possible to conduct an experiment before they become relevant. Secondly, we can employ higher density condensates and obtain a stronger Hawking signal, due to the reduced effect of loss on the mean-field.

The surface at which the normal component of the condensate flow exceeds the local speed of sound is termed {\it sonic horizon} \cite{visser:review}. As consequence of the causal disconnection of the supersonic and subsonic regions, quantum field theory predicts particle creation \cite{visser:essential}.
The \emph{analogue Hawking temperature} that characterizes their thermal spectrum is given by~\cite{visser:towards}
\begin{align}
T_{H}&=\frac{\hbar g_{h}}{2\pi k_{B} c_{h}},\:\:\:\:\:\:c_{h}=c(x_{h}),
\label{hawkingtemp1}
\\
g_{h}&=\frac{1}{2}\left| \hat{n}\cdot \nabla \left(c^{2} - v^{2} \right) \right|_{x=x_{h}} = c_{h}^{2} \left| \hat{n}\cdot \nabla  M \right|_{x=x_{h}}.
\label{surfacegrav}
\end{align}
Here $v$ is the flow speed, $c$ the speed of sound, $M=v/c$ the Mach number,  $x_{h}$ denotes a position on the horizon and $\hat{n}$ a normal vector to it. The underlying correspondence between the equations of motion for a scalar quantum field in curved space-time and for quantum phonons in a fluid can provide further analogies between cosmological effects and phenomena in a fluid~\cite{visser:review}. 

Values of $T_{H}$ for typical fluids are very small. Hence Bose-Einstein condensates have been considered as prime candidates for an observation of the analogue Hawking effect, owing to their low temperatures. We have shown in \rref{wuester:horizon} however, that to reach temperatures even of the order of $10$ nK, condensates have to be typically driven into a regime where three-body recombination significantly affects the \emph{mean-field} on a time scale of $50$ ms. Here we employ Bogoliubov theory used in \rref{dziarmaga:lossheating} to determine the consequences for the \emph{quantum field}.

%%%%%%%%%%%%%%%%%%%%%%%%%%%%%%%%
\ssection{Bogoliubov-de Gennes equations}
We split the field operator for bosonic atoms $\hat{\Psi}\ofx=\phi\ofx +\hat{\chi}\ofx$ into a condensate wavefunction $\phi\ofx=\langle \hat{\Psi}\ofx \rangle$ and its quantum fluctuations $\hat{\chi}\ofx$. The condensate wave function obeys the Gross-Pitaevskii equation
\begin{align}
i \hbar  \frac{\partial \phi}{\partial t}&={\cal L}\phi \equiv \left(-\frac{\hbar^2}{2m} \mathbf{\nabla}^{2} +W +g |\phi |^{2}\right)  \phi,
\label{3dgpe}
\end{align}
where $m$ is the atomic mass and $g$ the interaction strength related to the scattering length $a_{s}$ by $g =4 \pi \hbar^{2} a_{s}/m$, while $W$ denotes an external potential. We use $\int d^{3}\mathbf{x} |\phi(\mathbf{x})|^{2}=\sub{N}{cond}$, the number of condensate atoms. 
Where required, we use the notation $\rho=|\phi|^{2}$ for the condensate density, $v=i\hbar/(2m)[(\nabla \phi^{*}) \phi-\phi^{*}\nabla \phi]$ for its velocity and $c=\sqrt{g \rho/m}$ for the speed of sound. 

We decompose the fluctuating component as 
$
\hat{\chi}(\mathbf{x})=\sum_{n} [u_{n}(\mathbf{x})\hat{\alpha}_{n} + v_{n}^{*}(\mathbf{x})\hat{\alpha}^{\dagger}_{n}].
$
The presence of a subscript distinguishes references to the condensate velocity $v$ from those to the mode $v_{n}$.
The functions $u_{n}$, $v_{n}$ obey the Bogoliubov-de Gennes (BdG) equations~\cite{morgan:bdg,castin:lecture,piyush:nozzle}: $\sub{\cal L}{BdG}[u_{n}(\mathbf{x}),v_{n}(\mathbf{x})]^{T}=\epsilon_{n}[u_{n}(\mathbf{x}),v_{n}(\mathbf{x})]^{T}$, with
\begin{align}
&\sub{\cal L}{BdG}=
\left[ 
\begin{array}{cc}
{\cal L} -\mu +g \hat{Q} |\phi|^{2}\hat{Q} &  \!\!g  \hat{Q} \phi^{2}\hat{Q}^{*} \\
- g  \hat{Q}^{*} \phi^{* 2}\hat{Q} &\!\!\!\!\!\!\! -({\cal L} -\mu +g \hat{Q} |\phi|^{2}\hat{Q} )^{*}\\    
\end{array} 
\right],
\label{BdGoperator}
\end{align}
where $\hat{Q}=1 - |\phi\rangle\langle \phi|/\sub{N}{cond}$ projects onto the function space orthogonal to the condensate mode. The modes also must be normalized according to $\int d^{3}\mathbf{x}[|u_{n}(\mathbf{x})|^{2}-|v_{n}(\mathbf{x})|^{2}]=1$, for the fluctuations to obey bosonic commutation relations $[\hat{\alpha}_{n},\hat{\alpha}^{\dagger}_{m}]=\delta_{n,m}$.

For a homogeneous condensate in a quantization volume $V$ we have $W=0$, $\phi=\mu/g$. The BdG equations are then solved by $\uu{\mathbf{q}}{\mathbf{x}}=\frac{1}{\sqrt{V}}\tilde{u}_{\mathbf{q}}e^{i\mathbf{q} \mathbf{x}}$, $\tilde{u}_{\mathbf{q}}=\frac{\epsilon_{\mathbf{q}}+\epsilon_0}{2\sqrt{\epsilon_{\mathbf{q}}\epsilon_0}}$, $\vv{\mathbf{q}}{\mathbf{x}}=\frac{1}{\sqrt{V}}\tilde{v}_{\mathbf{q}}e^{i\mathbf{q}\mathbf{x}}$, $\tilde{v}_{\mathbf{q}}=-\frac{\epsilon_{\mathbf{q}}-\epsilon_0}{2\sqrt{\epsilon_{\mathbf{q}}\epsilon_0}}$, $\epsilon_0=\hbar^{2}\mathbf{q}^{2}/2m$, $\epsilon_{\mathbf{q}}=\sqrt{\hbar^{2}\mathbf{q}^{2}c^{2} + \epsilon_0^{2}}$~\cite{book:pethik,vogels:bogoliubov}.
%
%\begin{align}
%\uu{\mathbf{q}}{\mathbf{x}}&=\frac{1}{\sqrt{V}}\tilde{u}_{\mathbf{q}}e^{i\mathbf{q} \mathbf{x}}, \:\: \tilde{u}_{\mathbf{q}}=\frac{\epsilon_{\mathbf{q}}+\epsilon_0}{2\sqrt{\epsilon_{\mathbf{q}}\epsilon_0}},
%\CR
%\vv{\mathbf{q}}{\mathbf{x}}&=\frac{1}{\sqrt{V}}\tilde{v}_{\mathbf{q}}e^{i\mathbf{q}\mathbf{x}}, \:\: \tilde{v}_{\mathbf{q}}=-\frac{\epsilon_{\mathbf{q}}-\epsilon_0}{2\sqrt{\epsilon_{\mathbf{q}}\epsilon_0}},
%\CR
%\epsilon_0&=\hbar^{2}\mathbf{q}^{2}/2m, \:\: \epsilon_{\mathbf{q}}=\sqrt{\hbar^{2}\mathbf{q}^{2}c^{2} + \epsilon_0^{2} }.
%\label{homogenous_bogol_solutions}
%\end{align}
%
In the following we will assume that a sonic horizon is present in the condensate such that analogue Hawking radiation is created, but that the bulk condensate can still be considered as a homogeneous reservoir. No details regarding how to achieve this situation are required here, but can be found in \cite{wuester:horizon}.

%%%%%%%%%%%%%%%%%%%%%%%%%%%%%%%%
%%%%%%%%%%%%%%%%%%%%%%%%%%%%%%%%
\ssection{Loss induced phonons}
%%%%%%%%%%%%%%%%%%%%%%%%%%%%%%%%
Dziarmaga and Sacha have shown in \rref{dziarmaga:lossheating} that besides a reduction of the condensate population, atom losses also result in creation of phonons since the many-body quantum state is driven away from the Bogoliubov vacuum $|0\rangle$, defined by $\alpha_{n}|0\rangle=0$. These phonons will make a detection of analogue Hawking radiation more difficult. Thus the relative strength of the phonon sources has to be determined. 
We focus on three-body losses in what follows, since they are most prominent in BECs at high densities.

Three-body recombination results in a molecule and an energetic atom. The excess energy due to molecular binding, $E_{b}=\hbar^{2}/ma_{s}^{2}$, is split between the kinetic energies of molecule and fast atom in the ratio $1:2$. When these kinetic energies suffice for both particles to leave the trap, their corresponding quantum fields can be eliminated from the picture \cite{jack:loss}. One obtains an effective master equation that describes the effect of the loss process on the quantum state of the remaining trapped atoms. The generalization for $l$-body loss is~\cite{jack:loss,dziarmaga:lossheating}
\begin{align}
\frac{d\hat{\rho}}{dt}
&=\frac{1}{i\hbar} \comm{\hat{H}}{\hat{\rho}} 
+ \sum_{l}\gamma_l\intas {\cal{D} }[\hat{\Psi}^{l}]\hat{\rho},
\label{loss_mastereqn}
\end{align}
where ${\cal D}[\hat{a}]\hat{\rho}\equiv \hat{a}\hat{\rho}\hat{a}^{\dagger}-\hat{a}^{\dagger}\hat{a}\hat{\rho}/2-\hat{\rho}\hat{a}^{\dagger}\hat{a}/2$. $\hat{H}$ is the usual Hamiltonian describing the conservative dynamics of the remaining trapped atoms. Importantly, $\gamma_{l}$ is the \emph{event-rate} for a given loss process. Thus for example $\gamma_{3}=K_{3}/3$~\cite{ashton:loss}, where $K_{3}$ is the usual number loss rate for a condensate~\cite{footnote:K3def}.

Inserting the expansion of $\hat{\Psi}$, \eref{loss_mastereqn} can be rewritten in terms of quasi-particle operators $\hat{\alpha}_{m}$, $\hat{\alpha}^{\dagger}_{m}$. One then recognizes that each quasi-particle mode $m$ is coupled to a heat reservoir and its occupation $n_{m}(t)$ will relax towards a thermal state~\cite{dziarmaga:lossheating}. 
For the high density condensates that we consider, the three-body loss channel is strongly dominant~\cite{wuester:horizon}. If a single channel dominates, we can write~\cite{dziarmaga:lossheating}
\begin{align}
d n_{m}(t)/dt=-l \gamma_{l}\alpha_{lm}N^{l-1}(t)\left[n_{m}(t) -n_{lm}\right],
\label{lossheat_evolution}
\end{align}
where $N(t)$ is the number of condensate atoms. The coefficients $\alpha_{lm}$ and $n_{lm}$ are determined by~\cite{dziarmaga:lossheating}
\begin{align}
\intas |\phi_{0}|^{2(l-1)}|u_{m}|^{2}&=\alpha_{lm}(1+n_{lm}),
\CR
\intas |\phi_{0}|^{2(l-1)}|v_{m}|^{2}&=\alpha_{lm}n_{lm},
\label{coefficients1}
\end{align}
with the condensate mode $\phi_{0}$ defined by $\phi=\sqrt{N(t)}\phi_{0}$. \eref{lossheat_evolution} evolves each occupation number towards the equilibrium value $n_{lm}$. This value itself is time dependent in general, but can be assumed to vary slowly if losses are not too strong~\cite{dziarmaga:lossheating}.

In the following let the condensate be $d$-dimensional, with $D=3-d$ tightly confined transverse dimensions. We still allow $D=0$. The ideal setup for analogue Hawking radiation is not obvious: Elongated harmonically trapped condensates in 3D complicate matters with a nontrivial transverse structure of the sonic horizon~\cite{wuester:horizon}, whereas quasi 1D or 2D trapping must avoid a Tonks gas or quasi-condensate~\cite{petrov:tonks}. 
This could conflict with tightly confining the required high densities, as shown later.

In the longitudinal dimensions we imagine a homogenous condensate over a volume $V$. The three-dimensional vector $\mathbf{x}$ is decomposed as $\mathbf{x}=\mathbf{z}+\mathbf{r}_{\perp}$, where $\mathbf{z}$ is longitudinal and $\mathbf{r}_{\perp}$ transverse. With this splitting, we can write the condensate and Bogoliubov modes:
$\phi_{0}=A e^{-r_{\perp}^{2}/2\sigma^{2}}/\sqrt{V}$, 
$\uu{\mathbf{q}}{x}=\tilde{u}_{\mathbf{q}}e^{i\mathbf{q}\mathbf{z}}A e^{-r_{\perp}^{2}/2\sigma^{2}}/\sqrt{V}$, 
$\vv{\mathbf{q}}{x}=\tilde{v}_{\mathbf{q}}e^{i\mathbf{q}\mathbf{z}}A e^{-r_{\perp}^{2}/2\sigma^{2}}/\sqrt{V}$.
Here $\sigma$ is the ground state width of the transverse harmonic confinement and $r_{\perp}=|\mathbf{r}_{\perp}|$. We fix $A$ by $\int d^{D}\mathbf{r}_{\perp} A^{2} \exp{[-r_{\perp}^{2}\sigma^{2}]}=1$, which gives $A=\pi^{-(3-d)/4}\sigma^{-(3-d)/2}$. After replacing the discrete index $m$ by the continuous label $q$, \eref{coefficients1} can be solved by 
\begin{align}
n_{lq}&=|\tilde{v}_{q}|^{2},
\CR
\alpha_{lq}&=\pi^{(3-d)(1-l)/2}V^{1-L}l^{(d-3)/2}\sigma^{(3-d)(1-l)},
\label{coefficient_solutions}
\end{align}
using $|\tilde{u}_{q}|^{2}-|\tilde{v}_{q}|^{2}=1$. 

Let us denote the ``$l$-body loss temperature'' of this thermal state by $T_{Ll}$. We have $n_{lq}=(e^{\epsilon_{q}/k_{B}T_{Ll}} -1)^{-1}$, hence $k_{B}T_{Ll}=\epsilon_{q}/\log{[1/n_{lq}+1]}$. Since $n_{lq}=|\tilde{v}_{q}|^{2}\approx mc/2\hbar q\approx 1/\xi q \gg1$ for phonons, we can finally approximate
$k_{B}T_{Ll}=\epsilon_{q}n_{lq}$.
Using $\tilde{v}_{q}=\sqrt{mc/(2\hbar q)}$ and $\epsilon_{q}=\hbar qc$ for phononic wave numbers gives
\begin{align}
T_{Ll}=\frac{mc^{2}}{2k_{B}}.
\label{losstemperature2}
\end{align}
The Hawking temperature is limited to~\cite{wuester:horizon}
\begin{align}
T_{H}\lesssim\frac{mc^{2}}{\sqrt{2}\pi \Xi k_{B}}.
\label{hawkinglimit}
\end{align}
The factor $\Xi$ is defined by $|\nabla  M|_{x=x_{h}}=[\Xi \xi_{h}]^{-1}$, where $\xi_{h}$ is the healing length at the horizon~\cite{wuester:horizon}. We require $\Xi \gg1$ for hydrodynamic flow. We see that with both effects in equilibrium the loss-temperature is always greater than $T_{H}$. In a situation where atoms are continuously lost from the condensate, a rigorous equilibrium in which to interpret \eref{losstemperature2} does not exist. Nonetheless, when the loss is not too strong, we expect a quasi-equilibrium to apply. It is found that the actual heating can even slightly exceed \eref{losstemperature2}~\cite{dziarmaga:lossheating}.

%%%%%%%%%%%%%%%%%%%%%%%%%%%%%%%%
%%%%%%%%%%%%%%%%%%%%%%%%%%%%%%%%
\ssection{Heating time scale}
The equilibrium temperature associated with a loss process is independent of the corresponding loss (damping) rate, while the evolution towards equilibrium, \eref{lossheat_evolution}, is not~\cite{dziarmaga:lossheating}. A harmonic oscillator damped by a thermal bath behaves similar. Inserting the expression for $\alpha_{lm}$ one obtains
\begin{align}
\frac{d n_{m}(t)}{dt}=-l^{(d-1)/2}\tilde{\gamma}_{l}^{d}\sub{\rho}{cond}^{l-1}(t)\left[n_{m}(t) -n_{lm}\right],
\label{lossheat_evolution2}
\end{align}
where we use the condensate density in reduced dimensions $\sub{\rho}{cond} =N(t)/V$ and the effective loss rate
\begin{align}
\tilde{\gamma}_{l}^{d}=\pi^{(3-d)(1-l)/2}\sigma^{(3-d)(1-l)}\gamma_{l}.
\label{K3reduced}
\end{align}
If we consider short time scales on which $\sub{\rho}{cond}^{l-1}(t)$ can be treated as constant, the solution of \eref{lossheat_evolution2} is
\begin{align}
n_{m}(t)=n_{lm}\left[1-\exp{(-l^{(d-1)/2}\tilde{\gamma}_{l}^{d}\sub{\rho}{cond}^{l-1}t)}\right].
\label{phononevolution}
\end{align}
The time-scale on which the phonon population reaches its equilibrium value is therefore  $\tau=l^{(1-d)/2}\tilde{\gamma}_{l}^{-d}\sub{\rho}{cond}^{1-l}$. Since the reduced-dimensional density is related to the three-dimensional peak density by $\sub{\rho}{3D,peak}=\sub{\rho}{cond}\pi^{-(3-d)/2}\sigma^{-(3-d)}$, we can estimate $\tau$ using 3D quantities as
\begin{align}
\tau=l^{(1-d)/2}\gamma_{l}^{-1}\sub{\rho}{3D,peak}^{1-l}.
\label{lossheat_scale1}
\end{align}
For usual condensate densities $\tau$ can be quite large. One obtains $\tau=141$s for $^{23}$Na at $\sub{\rho}{3D}\sim 10^{20}$ m$^{-3}$ $(d=1)$. 
However, the densities required for reasonable analogue Hawking temperatures are significantly higher~\cite{wuester:horizon} with according to \eref{lossheat_evolution2} \emph{much} faster loss phonon creation. Let us parametrize the density as 
\begin{align}
\rho=\sqrt{3f/K_{3}\Delta t}, 
\label{densityparam}
\end{align}
which implies that within a time $\Delta t$ a fraction $f$ of this density will be lost due to three-body recombination~\cite{wuester:horizon}. We now \emph{choose} $\Delta t$ as the time which is required to detect the Hawking effect. We imply $\Delta t\sim 50$ ms in what follows, about the time required for quick spectroscopic phonon detection \cite{schuetzhold:phonondetection}.
\begin{table}
\begin{center}
\begin{tabular}{|r|cccc|}
\cline{1-5}
\Big. 
atom      & $^{4}$He      & $^{23}$Na  &$^{87}$Rb  & $^{137}$Cs   \\
\cline{1-5}
$\tabvspace g\times10^{50}$ [Jm$^3$]     & $15.7$    & $1$   & $0.5$ & $0.66$ \\
$\tabvspace K_{3}\times10^{42}$ [m$^{6}/$s]  & $9000$   & $2.12$  & $32$  & $130$ \\
$\tabvspace \sub{\rho}{max} \times 10^{-19}$ [m$^{-3}$]      & $3.0$  & $194$  & $50$  &  $25$ \\
$\tabvspace T_H$ [nK]  & $3.8$  & $16$  &  $2.2$  &  $1.4$  \\
\cline{1-5}
$\tabvspace \Big.\eta_{0}$ [\%] & 0.7 & 13 & 0.2 & 0.09\\   
\cline{1-5}
$\tabvspace m\sub{c}{max}^{2}/k_{B}$ [$\mu$K]  &  $0.34$ & $1.4$ & $0.2$ &  $0.25$ \\
$\tabvspace \hbar \omega_{\perp}/k_{B}$ [$\mu$K] & $1.2$ & $3$ & $0.4$ &  $0.5$\\
$\tabvspace \omega_{\perp}/2\pi$ [kHz] & $25$ & $125$ & $16.6$ & $10.4$ \\
$\tabvspace E_{b}/k_{B}$ [$\mu$K]   & $2100$ &  $2700$ & $190$ & $29$ \\
$\tabvspace E_{T}/k_{B}$ [$\mu$K] & $9.5$ & $22$  & $2.1$ & $0.68$ \\
\cline{1-5}
\end{tabular}
\end{center}
\caption{Comparison of common BEC species regarding analogue Hawking radiation and loss heating. The upper $4$ rows are reproduced from \rref{wuester:horizon}. 
Nextly we indicate $\eta_{0}$, defined as fraction of the original three-body loss rate that allows $T_{H}=30 nK$ for $\Delta t=50$ ms, while delaying the time scale for loss phonon production to $\tau\sim 20 \Delta t$, see \eref{maxlossfrac}. 
The lower $5$ rows show the hierarchy of energy scales $m c^{2}\ll \hbar \omega_{\perp}\ll \mbox{min}(E_{b},E_{T})$ required to achieve weakly interacting quasi 1D or 2D trapping, while still allowing three-body loss products to escape. The $\omega_{\perp}$ are selected to allow this. The mean field energies use $\rho=\sub{\rho}{max}$.
\label{atomcomparison}}
\end{table}
We arrive at a heating time scale under these conditions of
\begin{align}
\tau=\Delta t/3 f.
\label{lossheat_scale2}
\end{align}
This is of the order of the proposed measurement time $\Delta t$ and hence too short, unless we have $f\ll1/3$. However as was found in~\cite{wuester:horizon}, in the regime of such small loss the Hawking temperatures become problematically low. Hence, we usually have $\tau\sim\Delta t$. Alternatively we can calculate the time to create one phonon in a given spectral region, by the Hawking effect or by the loss. For the same parameter regime as above, one finds that the time scales are comparable. 

%%%%%%%%%%%%%%%%%%%%%%%%%%%%%%%%
%%%%%%%%%%%%%%%%%%%%%%%%%%%%%%%%
\ssection{Suppression of three-body loss}
It has been proposed to inhibit three-body loss processes in BECs by periodically flipping the phase of the weakly bound molecular state that causes the loss~\cite{search:lossinhibit}. The phase flip can be achieved using repeated $2\pi$ pulses of laser light resonant on an electronic exited state transition of the bound state. Destructive interference is then responsible for a reduction in three-body loss rates to only a few percent of their usual value. We now investigate whether this is sufficient to overcome three-body loss related obstacles to the creation of analogue Hawking radiation in BECs.

By \eref{lossheat_scale2}, we require a loss suppression that enables $f\ll1/3$. We pick a specified target temperature and assume three-body loss was reduced to $\tilde{K}_{3}=\eta K_{3}$, with $0<\eta<1$. The fractional loss within the measurement time $\Delta t$ will then be
\begin{align}
f=\left[2\pi^{2}(k_{B}T_{H})^{2} \Xi^{2}K_{3}\Delta t/3g^{2}\right] \eta\equiv f_{0} \eta.
\label{maxlossfrac}
\end{align}
For this we have eliminated $\rho$ between \eref{hawkinglimit} and \eref{densityparam} using $c=\sqrt{g \rho/m}$. See \tref{atomcomparison} for values of $\eta$ that allow analogue Hawking radiation at $T_{H}=30$ nK while separating the time-scales for phonon measurement and loss heating ($\tau\gg\Delta t$). 

A crucial indicator for the efficiency of the scheme presented in \cite{search:lossinhibit} is the number of laser pulses that fit into the average life-time of a molecule, before its quantum state is perturbed by a collision with a condensate atom. This life-time can be estimated as $\sub{\tau}{mol}=(\kappa \rho)^{-1}$, where $\kappa\sim10^{-16}m^{3}s^{-1}$ for $^{87}$Rb \cite{yurovsky:kapparb} and $^{23}$Na \cite{yurovsky:kappana}. For densities $\sub{\rho}{Rb}=5\times10^{20}m^{-3}$, $\sub{\rho}{Na}=2\times10^{21}m^{-3}$, we obtain $\sub{\tau}{mol}\sim10 \mu$s. According to \rref{search:lossinhibit} this allows a loss reduction to a few percent.

A side effect of the laser pulses that suppress three-body losses is one-body loss due to Rayleigh scattering of laser photons \cite{search:lossinhibit}. The increased one-body loss rate has been estimated as $\gamma_{1}\sim0.1$ s$^{-1}$. We then see from \eref{lossheat_scale1} that the time scale for one-body loss induced phonon creation is about $10$ s and hence unproblematic.

%%%%%%%%%%%%%%%%%%%%%%%%%%%%%%%%
%%%%%%%%%%%%%%%%%%%%%%%%%%%%%%%%
\ssection{Reaction products of loss process}
%%%
Given the importance of three-body losses, we have to address the evolution of the molecules and fast atoms created in the recombination process. For \eref{loss_mastereqn} to be valid, it is required that they are energetic enough to leave the trap \cite{jack:loss}. Also for the cosmological analogy to hold, we wish to avoid the complications of a coupled atom-molecular condensate. Finally, collisions between the loss products and remaining atoms would induce further unwanted heating if the loss products remained in the trap~\cite{grimm:largea}.

To ensure that the molecules and fast atoms can leave the trap, we require the trap depth characterized by $\hbar \omega_{\perp}$ to satisfy $\hbar \omega_{\perp} \ll E_{b}$. To avoid trapping so tight that we enter the Tonks gas regime, the parameter $\gamma=[\sub{n}{1d}\xi_{1d}]^{-2}$ as to be much smaller than one~\cite{petrov:tonks}. Here $\sub{n}{1d}=\pi \rho \sigma^2$ and $\xi_{1d}= \hbar/\sqrt{m \sub{n}{1d} \sub{g}{1d}}$ are the 1D density and healing length respectively, with $\sigma=(\hbar/m\omega_{\perp})^{-1/2}$ and $\sub{g}{1d}=g/(2\pi\sigma^2)$. The condition can be reformulated as $\hbar \omega_{\perp} \ll E_{T}\equiv \hbar^2 \sqrt{\pi \rho/2 a_{s}}/m$. If we wish to study the sonic horizon in a quasi one or two dimensional setup, the strength of transverse confinement is constrained by $m c^{2}\ll \hbar \omega_{\perp}\ll \mbox{min}(E_{b},E_{T})$, where $m c^{2}$ is the interaction energy of the confined condensate. 
Exemplary numbers for these energies are shown in \tref{atomcomparison}, which demonstrate that this hierarchy can usually only just be fulfilled.

%%%%%%%%%%%%%%%%%%%%%%%%%%%%%%%%
\ssection{Conclusions}
We have shown that loss induced phonons are an overwhelming background for analogue Hawking radiation in a Bose-Einstein condensate. To overcome this problem we suggest a moderate suppression of three-body losses. This can make the time-scale of loss induced phonon creation sufficiently long for a fast detection of the analogue Hawking effect. 

%%%%%%%%%%%%%%%%%%%%%%%%%%%%%%%%
\acknowledgments
We thank R.~Sch\"utzhold for drawing loss heating to our attention and further fruitful discussions.
It is also a pleasure to acknowledge discussions with C.~Savage.


\begin{thebibliography}{31}
\expandafter\ifx\csname natexlab\endcsname\relax\def\natexlab#1{#1}\fi
\expandafter\ifx\csname bibnamefont\endcsname\relax
  \def\bibnamefont#1{#1}\fi
\expandafter\ifx\csname bibfnamefont\endcsname\relax
  \def\bibfnamefont#1{#1}\fi
\expandafter\ifx\csname citenamefont\endcsname\relax
  \def\citenamefont#1{#1}\fi
\expandafter\ifx\csname url\endcsname\relax
  \def\url#1{\texttt{#1}}\fi
\expandafter\ifx\csname urlprefix\endcsname\relax\def\urlprefix{URL }\fi
\providecommand{\bibinfo}[2]{#2}
\providecommand{\eprint}[2][]{\url{#2}}

\bibitem[{\citenamefont{Hawking}(1975)}]{hawking:hr1}
\bibinfo{author}{\bibfnamefont{S.~W.} \bibnamefont{Hawking}},
  \bibinfo{journal}{Commun. Math. Phys.} \textbf{\bibinfo{volume}{43}},
  \bibinfo{pages}{199} (\bibinfo{year}{1975}).

\bibitem[{\citenamefont{Hawking}(1974)}]{hawking:hr2}
\bibinfo{author}{\bibfnamefont{S.~W.} \bibnamefont{Hawking}},
  \bibinfo{journal}{Nature (London)} \textbf{\bibinfo{volume}{248}},
  \bibinfo{pages}{30} (\bibinfo{year}{1974}).

\bibitem[{\citenamefont{Unruh}(1981)}]{unruh:bholes}
\bibinfo{author}{\bibfnamefont{W.~G.} \bibnamefont{Unruh}},
  \bibinfo{journal}{Phys. Rev. Lett.} \textbf{\bibinfo{volume}{46}},
  \bibinfo{pages}{1351} (\bibinfo{year}{1981}).

\bibitem[{\citenamefont{Barcel{\'o} et~al.}(2005)\citenamefont{Barcel{\'o},
  Liberati, and Visser}}]{visser:review}
\bibinfo{author}{\bibfnamefont{C.}~\bibnamefont{Barcel{\'o}}},
  \bibinfo{author}{\bibfnamefont{S.}~\bibnamefont{Liberati}}, \bibnamefont{and}
  \bibinfo{author}{\bibfnamefont{M.}~\bibnamefont{Visser}},
  \bibinfo{journal}{Living Rev. Relativity} \textbf{\bibinfo{volume}{8}},
  \bibinfo{pages}{12} (\bibinfo{year}{2005}).

\bibitem[{\citenamefont{Barcel{\'o} et~al.}(2003)\citenamefont{Barcel{\'o},
  Liberati, and Visser}}]{visser:towards}
\bibinfo{author}{\bibfnamefont{C.}~\bibnamefont{Barcel{\'o}}},
  \bibinfo{author}{\bibfnamefont{S.}~\bibnamefont{Liberati}}, \bibnamefont{and}
  \bibinfo{author}{\bibfnamefont{M.}~\bibnamefont{Visser}},
  \bibinfo{journal}{Int. J. Mod. Phys. A} \textbf{\bibinfo{volume}{18}},
  \bibinfo{pages}{3735} (\bibinfo{year}{2003}).

\bibitem[{\citenamefont{Garay et~al.}(2000)\citenamefont{Garay, Anglin, Cirac,
  and Zoller}}]{garay:prl}
\bibinfo{author}{\bibfnamefont{L.~J.} \bibnamefont{Garay}},
  \bibinfo{author}{\bibfnamefont{J.~R.} \bibnamefont{Anglin}},
  \bibinfo{author}{\bibfnamefont{J.~I.} \bibnamefont{Cirac}}, \bibnamefont{and}
  \bibinfo{author}{\bibfnamefont{P.}~\bibnamefont{Zoller}},
  \bibinfo{journal}{Phys. Rev. Lett.} \textbf{\bibinfo{volume}{85}},
  \bibinfo{pages}{4643} (\bibinfo{year}{2000}).

\bibitem[{\citenamefont{Garay et~al.}(2001)\citenamefont{Garay, Anglin, Cirac,
  and Zoller}}]{garay:pra}
\bibinfo{author}{\bibfnamefont{L.~J.} \bibnamefont{Garay}},
  \bibinfo{author}{\bibfnamefont{J.~R.} \bibnamefont{Anglin}},
  \bibinfo{author}{\bibfnamefont{J.~I.} \bibnamefont{Cirac}}, \bibnamefont{and}
  \bibinfo{author}{\bibfnamefont{P.}~\bibnamefont{Zoller}},
  \bibinfo{journal}{Phys. Rev. A} \textbf{\bibinfo{volume}{63}},
  \bibinfo{pages}{023611} (\bibinfo{year}{2001}).

\bibitem[{\citenamefont{Giovanazzi et~al.}(2004)\citenamefont{Giovanazzi,
  Farrell, Kiss, and Leonhardt}}]{giovanazzi:horizon}
\bibinfo{author}{\bibfnamefont{S.}~\bibnamefont{Giovanazzi}},
  \bibinfo{author}{\bibfnamefont{C.}~\bibnamefont{Farrell}},
  \bibinfo{author}{\bibfnamefont{T.}~\bibnamefont{Kiss}}, \bibnamefont{and}
  \bibinfo{author}{\bibfnamefont{U.}~\bibnamefont{Leonhardt}},
  \bibinfo{journal}{Phys. Rev. A} \textbf{\bibinfo{volume}{70}},
  \bibinfo{pages}{063602} (\bibinfo{year}{2004}).

\bibitem[{\citenamefont{Barcel{\'o} et~al.}(2001)\citenamefont{Barcel{\'o},
  Liberati, and Visser}}]{barcelo:diffmetric}
\bibinfo{author}{\bibfnamefont{C.}~\bibnamefont{Barcel{\'o}}},
  \bibinfo{author}{\bibfnamefont{S.}~\bibnamefont{Liberati}}, \bibnamefont{and}
  \bibinfo{author}{\bibfnamefont{M.}~\bibnamefont{Visser}},
  \bibinfo{journal}{Class. Quant. Grav.} \textbf{\bibinfo{volume}{18}},
  \bibinfo{pages}{1137} (\bibinfo{year}{2001}).

\bibitem[{\citenamefont{Carusotto et~al.}(2008)\citenamefont{Carusotto,
  Fagnocchi, Recati, Balbinot, and Fabbri}}]{fabbri:wigner}
\bibinfo{author}{\bibfnamefont{I.}~\bibnamefont{Carusotto}},
  \bibinfo{author}{\bibfnamefont{S.}~\bibnamefont{Fagnocchi}},
  \bibinfo{author}{\bibfnamefont{A.}~\bibnamefont{Recati}},
  \bibinfo{author}{\bibfnamefont{R.}~\bibnamefont{Balbinot}}, \bibnamefont{and}
  \bibinfo{author}{\bibfnamefont{A.}~\bibnamefont{Fabbri}}
  (\bibinfo{year}{2008}), \eprint{cond-mat.other/0803.0507}.

\bibitem[{\citenamefont{W{\"u}ster and Savage}(2007)}]{wuester:horizon}
\bibinfo{author}{\bibfnamefont{S.}~\bibnamefont{W{\"u}ster}} \bibnamefont{and}
  \bibinfo{author}{\bibfnamefont{C.~M.} \bibnamefont{Savage}},
  \bibinfo{journal}{Phys. Rev. A} \textbf{\bibinfo{volume}{76}},
  \bibinfo{pages}{013608} (\bibinfo{year}{2007}).

\bibitem[{\citenamefont{Sch{\"u}tzhold}(2008)}]{schuetzhold:rsoc}
\bibinfo{author}{\bibfnamefont{R.}~\bibnamefont{Sch{\"u}tzhold}}
  (\bibinfo{year}{2008}), \bibinfo{note}{to appear in Phil.\ Trans.\ Roy.\
  Soc.\ (London) A (2008)}.

\bibitem[{\citenamefont{Dziarmaga and Sacha}(2003)}]{dziarmaga:lossheating}
\bibinfo{author}{\bibfnamefont{J.}~\bibnamefont{Dziarmaga}} \bibnamefont{and}
  \bibinfo{author}{\bibfnamefont{K.}~\bibnamefont{Sacha}},
  \bibinfo{journal}{Phys. Rev. A} \textbf{\bibinfo{volume}{68}},
  \bibinfo{pages}{043607} (\bibinfo{year}{2003}).

\bibitem[{foo({\natexlab{a}})}]{footnote:correlations}
\bibinfo{note}{Correlations between excitations on either side of the sonic
  black hole can in principle distinguish between analogue Hawking phonons and
  those due to loss heating \cite{fabbri:corrfct,fabbri:wigner}, however the
  thermal fluctuations due to the latter will still reduce the correlation
  signal.}

\bibitem[{\citenamefont{Search et~al.}(2004)\citenamefont{Search, Zhang, and
  Meystre}}]{search:lossinhibit}
\bibinfo{author}{\bibfnamefont{C.~P.} \bibnamefont{Search}},
  \bibinfo{author}{\bibfnamefont{W.}~\bibnamefont{Zhang}}, \bibnamefont{and}
  \bibinfo{author}{\bibfnamefont{P.}~\bibnamefont{Meystre}},
  \bibinfo{journal}{Phys. Rev. Lett.} \textbf{\bibinfo{volume}{92}},
  \bibinfo{pages}{140401} (\bibinfo{year}{2004}).

\bibitem[{\citenamefont{Yurovsky and Band}(2007)}]{yurovsky:lessloss}
\bibinfo{author}{\bibfnamefont{V.~A.} \bibnamefont{Yurovsky}} \bibnamefont{and}
  \bibinfo{author}{\bibfnamefont{Y.~B.} \bibnamefont{Band}},
  \bibinfo{journal}{Phys. Rev. A} \textbf{\bibinfo{volume}{75}},
  \bibinfo{pages}{012717} (\bibinfo{year}{2007}).

\bibitem[{\citenamefont{Visser}(2003)}]{visser:essential}
\bibinfo{author}{\bibfnamefont{M.}~\bibnamefont{Visser}},
  \bibinfo{journal}{Int. J. Mod. Phys. D} \textbf{\bibinfo{volume}{12}},
  \bibinfo{pages}{649} (\bibinfo{year}{2003}).

\bibitem[{\citenamefont{Morgan et~al.}(1998)\citenamefont{Morgan, Choi,
  Burnett, and Edwards}}]{morgan:bdg}
\bibinfo{author}{\bibfnamefont{S.~A.} \bibnamefont{Morgan}},
  \bibinfo{author}{\bibfnamefont{S.}~\bibnamefont{Choi}},
  \bibinfo{author}{\bibfnamefont{K.}~\bibnamefont{Burnett}}, \bibnamefont{and}
  \bibinfo{author}{\bibfnamefont{M.}~\bibnamefont{Edwards}},
  \bibinfo{journal}{Phys. Rev. A} \textbf{\bibinfo{volume}{57}},
  \bibinfo{pages}{3818} (\bibinfo{year}{1998}).

\bibitem[{\citenamefont{Castin}(2001)}]{castin:lecture}
\bibinfo{author}{\bibfnamefont{Y.}~\bibnamefont{Castin}}, in
  \emph{\bibinfo{booktitle}{Ecole d'Ete de Physique Theorique}}, edited by
  \bibinfo{editor}{\bibfnamefont{R.}~\bibnamefont{Kaiser}},
  \bibinfo{editor}{\bibfnamefont{C.}~\bibnamefont{Westbrook}},
  \bibnamefont{and} \bibinfo{editor}{\bibfnamefont{F.}~\bibnamefont{David}}
  (\bibinfo{publisher}{EDP Sciences and Springer-Verlag},
  \bibinfo{address}{Berlin}, \bibinfo{year}{2001}), vol.~\bibinfo{volume}{72}.

\bibitem[{\citenamefont{Jain et~al.}(2007)\citenamefont{Jain, Bradley, and
  Gardiner}}]{piyush:nozzle}
\bibinfo{author}{\bibfnamefont{P.}~\bibnamefont{Jain}},
  \bibinfo{author}{\bibfnamefont{A.~S.} \bibnamefont{Bradley}},
  \bibnamefont{and} \bibinfo{author}{\bibfnamefont{C.~W.}
  \bibnamefont{Gardiner}}, \bibinfo{journal}{Phys. Rev. A}
  \textbf{\bibinfo{volume}{76}}, \bibinfo{pages}{023617}
  (\bibinfo{year}{2007}).

\bibitem[{\citenamefont{Pethik and Smith}(2002)}]{book:pethik}
\bibinfo{author}{\bibfnamefont{C.~J.} \bibnamefont{Pethik}} \bibnamefont{and}
  \bibinfo{author}{\bibfnamefont{H.}~\bibnamefont{Smith}},
  \emph{\bibinfo{title}{Bose-Einstein condensation in dilute gases}}
  (\bibinfo{publisher}{Cambridge University Press}, \bibinfo{year}{2002}).

\bibitem[{\citenamefont{Vogels et~al.}(2002)\citenamefont{Vogels, Xu, Raman,
  {Abo-Shaeer}, and Ketterle}}]{vogels:bogoliubov}
\bibinfo{author}{\bibfnamefont{J.~M.} \bibnamefont{Vogels}},
  \bibinfo{author}{\bibfnamefont{K.}~\bibnamefont{Xu}},
  \bibinfo{author}{\bibfnamefont{C.}~\bibnamefont{Raman}},
  \bibinfo{author}{\bibfnamefont{J.~R.} \bibnamefont{{Abo-Shaeer}}},
  \bibnamefont{and} \bibinfo{author}{\bibfnamefont{W.}~\bibnamefont{Ketterle}},
  \bibinfo{journal}{Phys. Rev. Lett.} \textbf{\bibinfo{volume}{88}},
  \bibinfo{pages}{060402} (\bibinfo{year}{2002}).

\bibitem[{\citenamefont{Jack}(2002)}]{jack:loss}
\bibinfo{author}{\bibfnamefont{M.~W.} \bibnamefont{Jack}},
  \bibinfo{journal}{Phys. Rev. Lett.} \textbf{\bibinfo{volume}{89}},
  \bibinfo{pages}{140402} (\bibinfo{year}{2002}).

\bibitem[{\citenamefont{Norrie et~al.}(2006)\citenamefont{Norrie, Ballagh,
  Gardiner, and Bradley}}]{ashton:loss}
\bibinfo{author}{\bibfnamefont{A.~A.} \bibnamefont{Norrie}},
  \bibinfo{author}{\bibfnamefont{R.~J.} \bibnamefont{Ballagh}},
  \bibinfo{author}{\bibfnamefont{C.~W.} \bibnamefont{Gardiner}},
  \bibnamefont{and} \bibinfo{author}{\bibfnamefont{A.~S.}
  \bibnamefont{Bradley}}, \bibinfo{journal}{Phys. Rev. A}
  \textbf{\bibinfo{volume}{73}}, \bibinfo{pages}{043618}
  (\bibinfo{year}{2006}).

\bibitem[{foo({\natexlab{b}})}]{footnote:K3def}
\bibinfo{note}{For this loss rate $dN(t)/dt=-K_{3}\int d^{3}x n(x)^{3}$ holds.}

\bibitem[{\citenamefont{Petrov et~al.}(2000)\citenamefont{Petrov, Shlyapnikov,
  and Walraven}}]{petrov:tonks}
\bibinfo{author}{\bibfnamefont{D.~S.} \bibnamefont{Petrov}},
  \bibinfo{author}{\bibfnamefont{G.~V.} \bibnamefont{Shlyapnikov}},
  \bibnamefont{and} \bibinfo{author}{\bibfnamefont{J.~T.~M.}
  \bibnamefont{Walraven}}, \bibinfo{journal}{Phys. Rev. Lett.}
  \textbf{\bibinfo{volume}{85}}, \bibinfo{pages}{3745} (\bibinfo{year}{2000}).

\bibitem[{\citenamefont{Sch{\"u}tzhold}(2006)}]{schuetzhold:phonondetection}
\bibinfo{author}{\bibfnamefont{R.}~\bibnamefont{Sch{\"u}tzhold}},
  \bibinfo{journal}{Phys. Rev. Lett.} \textbf{\bibinfo{volume}{97}},
  \bibinfo{pages}{190405} (\bibinfo{year}{2006}).

\bibitem[{\citenamefont{Yurovsky and {Ben-Reuven}}(2003)}]{yurovsky:kapparb}
\bibinfo{author}{\bibfnamefont{V.~A.} \bibnamefont{Yurovsky}} \bibnamefont{and}
  \bibinfo{author}{\bibfnamefont{A.}~\bibnamefont{{Ben-Reuven}}},
  \bibinfo{journal}{Phys. Rev. A} \textbf{\bibinfo{volume}{67}},
  \bibinfo{pages}{050701(R)} (\bibinfo{year}{2003}).

\bibitem[{\citenamefont{Yurovsky et~al.}(1999)\citenamefont{Yurovsky,
  {Ben-Reuven}, Julienne, and Williams}}]{yurovsky:kappana}
\bibinfo{author}{\bibfnamefont{V.~A.} \bibnamefont{Yurovsky}},
  \bibinfo{author}{\bibfnamefont{A.}~\bibnamefont{{Ben-Reuven}}},
  \bibinfo{author}{\bibfnamefont{P.~S.} \bibnamefont{Julienne}},
  \bibnamefont{and} \bibinfo{author}{\bibfnamefont{C.~J.}
  \bibnamefont{Williams}}, \bibinfo{journal}{Phys. Rev. A}
  \textbf{\bibinfo{volume}{60}}, \bibinfo{pages}{R765} (\bibinfo{year}{1999}).

\bibitem[{\citenamefont{Weber et~al.}(2003)\citenamefont{Weber, Herbig, Mark,
  N{\"a}gerl, and Grimm}}]{grimm:largea}
\bibinfo{author}{\bibfnamefont{T.}~\bibnamefont{Weber}},
  \bibinfo{author}{\bibfnamefont{J.}~\bibnamefont{Herbig}},
  \bibinfo{author}{\bibfnamefont{M.}~\bibnamefont{Mark}},
  \bibinfo{author}{\bibfnamefont{H.-C.} \bibnamefont{N{\"a}gerl}},
  \bibnamefont{and} \bibinfo{author}{\bibfnamefont{R.}~\bibnamefont{Grimm}},
  \bibinfo{journal}{Phys. Rev. Lett.} \textbf{\bibinfo{volume}{91}},
  \bibinfo{pages}{123201} (\bibinfo{year}{2003}).

\bibitem[{\citenamefont{Balbinot et~al.}(2007)\citenamefont{Balbinot, Fabbri,
  and Fagnocchi}}]{fabbri:corrfct}
\bibinfo{author}{\bibfnamefont{R.}~\bibnamefont{Balbinot}},
  \bibinfo{author}{\bibfnamefont{A.}~\bibnamefont{Fabbri}}, \bibnamefont{and}
  \bibinfo{author}{\bibfnamefont{S.}~\bibnamefont{Fagnocchi}}
  (\bibinfo{year}{2007}), \eprint{cond-mat.other/0711.4520}.

\end{thebibliography}
\end{document}